\documentclass[preprint,aps,nofootinbib]{revtex4}
\usepackage{}
\usepackage{graphicx}
\usepackage{amsmath}
\usepackage{amsfonts}
\usepackage{amssymb}
\usepackage{color}%
\usepackage{dcolumn}
\providecommand{\U}[1]{\protect\rule{.1in}{.1in}}

\newcommand{\f}{\begin{equation}}
\newcommand{\ff}{\end{equation}}
\newcommand{\fa}{\begin{eqnarray}}
\newcommand{\ffa}{\end{eqnarray}}

\begin{document}
\title{From Petrov-Einstein to Navier-Stokes in Spatially Curved Spacetime }
\author{Tai-Zhuo Huang $^{1}$}
\email{huangtaizhuo@126.com}
\author{Yi Ling $^{1,2}$}
\email{yling@ncu.edu.cn}
\author{Wen-Jian Pan $^{1}$}
\email{wjpan_zhgkxy@163.com}
\author{Yu Tian $^{3}$}
\email{ytian@gucas.ac.cn}
\author{Xiao-Ning Wu $^{2}$} \email{wuxn@amss.ac.cn}

\affiliation{$^1$Center for Relativistic Astrophysics and High
Energy Physics, Department of Physics, Nanchang University, 330031,
China\\$^2$ Institute of Mathematics, Academy of Mathematics and
System Science, CAS, Beijing 100190, China and Hua Loo-Keng Key
Laboratory of Mathematics, CAS, Beijing 100190, China \\$^3$ College
of Physical Sciences, Graduate University of Chinese Academy of
Sciences, Beijing 100049, China}

\begin{abstract}
We generalize the framework in arXiv:1104.5502 to the case that an
embedding may have a non-vanishing intrinsic curvature. Directly
employing the Brown-York stress tensor as the fundamental variables,
we study the effect of finite perturbations of the extrinsic
curvature while keeping the intrinsic metric fixed. We show that
imposing a Petrov type I condition on the hypersurface geometry may
reduce to the incompressible Navier-Stokes equation for a fluid
moving in spatially curved spacetime in the near-horizon limit.
\end{abstract} \maketitle

\section {Introduction}
It has been more than three decades since Damour firstly found that
the excitations of a black hole horizon behave very much like those
of a fluid\cite{Damour1979,P-T}. Later the analogy of gravity
dynamics with hydrodynamics has been further disclosed in
\cite{Jacobson:1995ab} and then in the context of AdS/CFT
correspondence\cite{PSS,KSS,B-L,I-L,Bhattacharyya:2008kq,EFO}. In
particular it is shown in \cite{Bredberg:2011jq} that in a suitably
defined near horizon limit, the dynamics of gravity on an arbitrary
cutoff surface can be governed by the incompressible Navier-Stokes
equation if one identifies the Brown-York tensor of the bulk gravity
with the stress-energy tensor of a fluid moving on the surface(and
other relevant work can be found, for instance, in
\cite{Padmanabhan:2010rp,Bredberg:2010ky,Compere:2011dx,Cai,NTWL}).
More specifically, in this approach one embeds a $p+1$-dimensional
timelike hypersurface into a $p+2$-dimensional spacetime, with a
small distance from the horizon. Then imposing a Dirichlet-like
boundary condition on the hypersurface while demanding regularity on
the future horizon and no incoming flux across the past horizon, one
can solve the bulk Einstein equations explicitly under the
non-relativistical limit with a long-wavelength expansion, and the
incompressible Navier-Stokes equation can be obtained as a
constraint on the hypersurface.

Very recently it has been remarkably noted in \cite{Lysov:2011xx}
that imposing a Petrov type I condition on the hypersurface geometry
exactly reduces the degrees of freedom in the extrinsic curvature to
those of a fluid such that the leading-order Einstein constraint
leads to the Navier-Stokes equation provided that the mean curvature
of the embedding is large enough. This observation strongly implies
that regularity on the future horizon and the Petrov type I
condition are equivalent at least in the near horizon limit.
However, mathematically imposing the Petrov condition is much
simpler and elegant than imposing regularity. It is very worthy to
further understand its role in exploring the deep relations between
the Einstein equations and Navier-Stokes equations.

Since in \cite{Lysov:2011xx} only an intrinsically flat
$p+1$-dimensional embedding is taken into account, in this paper we
intend to generalize this framework to the case that the embedded
hypersurface has a non-vanishing intrinsic curvature. Through
explicit construction we will show that for a spatially curved
embedding, it is still possible to obtain an incompressible
Navier-Stokes equation for a fluid moving in this background. We
organize the paper as follows. In next section we present the
generalized framework in which the embedded hypersurface is
intrinsically curved. In section three and four we explicitly
construct two models in which the background metric is spatially
curved, and then study the fluctuations of the extrinsic curvature
on the hypersurface. Imposing the Petrov type I condition as well as
the Hamiltonian constraint, we obtain the Navier-Stokes equations
with incompressible condition in the near horizon limit with a large
mean curvature. Moreover, in contrast to the scheme used in
\cite{Lysov:2011xx} where a new traceless tress tensor is
introduced, we insist to expand the effect of fluctuations directly
in terms of the Brown-York stress tensor. To demonstrate that our
scheme derives the same results at least at the leading orders of
the expansion as in \cite{Lysov:2011xx}, we present two simple
examples which has a Minkowski limit in the appendix.

\section {The framework for an intrinsically curved embedding}
Given a $p+2$-dimensional spacetime with a bulk metric $g_{\mu\nu}$
which satisfies the vacuum Einstein's equations
\begin{equation}
    G_{\mu\nu}=0,  \ \ \ \mu, \nu =0,..., p+1.
\end{equation}
We consider an embedding $\Sigma_c$ whose $p+1$-dimensional
spacetime with an induced metric $\gamma_{ab}$ may be intrinsically
curved. Suppose that the extrinsic curvature of the hypersurface is
$K_{ab}$, then the $p+1$ ``momentum constraints'' on $\Sigma_c$
reads as
\begin{equation}
    \nabla^a(K_{ab}-\gamma_{ab}K)=0,   \ \ \ \  a,b=0,..., p
\end{equation}
where $\nabla_a$ is compatible with the induced metric on
$\Sigma_c$, namely $\nabla_a\gamma_{bc}=0$. While the ``Hamiltonian
constraint'' is
\begin{equation}
    ^{p+1}R+K_{ab}K^{ab}-K^2=0.\label{HC}
\end{equation}
When the bulk metric satisfies the vacuum Einstein equation, the
Riemann curvature tensor and the Weyl tensor are equal, thus we can
decompose the Weyl tensor in $p+2$ dimensions in terms of the
intrinsic curvature of the $p+1$ hypersurface and its extrinsic
curvature. It turns out that the result is
\begin{eqnarray}
    &&C_{abcd} = {}^{p+1}R_{abcd} + K_{ad}K_{bc} - K_{ac}K_{bd} \nonumber\\
    &&C_{abc(n)} = \nabla_aK_{bc} - \nabla_bK_{ac} \nonumber\\
    &&C_{a(n)b(n)}=- \ {^{p+1}R_{ab}} + KK_{ab}
    -K_{ac}{K^c}_b,\label{Weyl}
\end{eqnarray}
where $C_{abc(n)}\equiv C_{abc\mu}n^{\mu}$ and $n^\mu$ is the unit
normal to $\Sigma_c$. The Petrov type I condition is defined as
\begin{equation}
    C_{(\ell)i(\ell)j} = \ell^\mu {m_i}^\nu \ell^\alpha {m_j}^\beta
    C_{\mu\nu\alpha\beta} = 0,
\end{equation}
where $p+2$ Newman-Penrose-like vector fields should satisfy the
relations
\begin{equation}
    \ell^2 = k^2=0, \ \ \ \ \ \ \ \ (k,\ell)=1,\ \ \ \ \ \  (k,m_i)=(\ell,m_i)=0,\ \ \ \ \ \
    (m^i,m_j)={\delta^i}_j.
\end{equation}

\section{The Petrov type I conditions for spatially curved embedding}
In this section we consider a $p+2$ dimensional space time with a
metric as
\begin{equation}
    {ds^2}_{p+2} = -rdt^2 + 2dtdr +
h_{ij}(x^i)dx^idx^j,\label{back}
\end{equation}
where $h_{ij}(x^i)$ is a general spatial metric but independent of
the coordinates $t$ and $r$.

We consider an embedding $\Sigma_c$ by setting $r=r_{c}$ such that
the induced metric $\gamma_{ab}$ on $\Sigma_c$ is
\begin{equation}
 {ds^2}_{p+1}=-r_{c}dt^2+h_{ij}(x^i)dx^i dx^j\equiv -(dx^0)^2+h_{ij}(x^i)dx^i dx^j.
\end{equation}
We also require that the induced metric on $\Sigma_c$ is fixed and
then only consider the effects of the fluctuations of the
extrinsic curvature. Now it is straightforward to obtain the
components of $K_{ab}$ as
\begin{eqnarray}
    K_{tt} &=& -{1\over{2}}\sqrt{r_c},\ \ \ \ \ \ K_{ti} = \ 0,\nonumber\\
    K_{ij} &=& \,0, \ \ \ \ \ \ \ \ \ \ \ \ \ \ K \ =
    {1\over{2\sqrt{r_c}}},
\end{eqnarray}
where $K$ is the trace of the extrinsic curvature. Equivalently we
may define the Brown-York stress tensor on $\Sigma_c$ as
\begin{equation}
    t_{ab}\equiv K\gamma_{ab}-K_{ab}.
\end{equation}
Next we introduce a parameter $\lambda$ by rescaling the time
coordinate with $\tau = \lambda x^0$  in order to discuss the
dynamical behavior of the geometry in the non-relativistical limit,
i.e.
\begin{equation}
 {ds^2}_{p+1}={-{1\over{\lambda^2}}d\tau^2+h_{ij}(x^i)dx^idx^j}.\label{induced}
\end{equation}
Moreover, we identify the parameter $\lambda$ with the location of
the hypersurface by setting $r_c=\lambda^2$ such that the limit
$\lambda \rightarrow 0$ means a large mean curvature and can be
thought of as a kind of near-horizon limit. In this coordinate
system we obtain the relations between the Brown-York stress tensor
and the extrinsic curvature as follows
\begin{eqnarray}
    {K^\tau}_\tau &=&  K-{t^\tau}_\tau={t\over{p}}-{t^\tau}_\tau,\ \ \ \ \ \ \ \ {K^\tau}_i = -{t^\tau}_i, \nonumber\\
    {K^i}_j       &=&  -{t^i}_j+{\delta^i}_j{t\over{p}},\ \ \ \ \ \ \ \ \ \ \ \ \ \ \ \ \ \  K \,\,\,= \ {t\over{p}}.
\end{eqnarray}
It is easy to show that except $\Gamma^i_{jk}$, all the other
components of the connection with the induced metric (\ref{induced})
on $\Sigma_c$ vanish.

Furthermore, the requirement that the background (\ref{back})
should satisfy the Einstein vacuum equations in $p+2$-dimensional
spacetime leads to
\begin{equation}
    ^{p+1}R={}^{p}R=0, \ \ \ \ \ \ \ ^{p+1}R_{ij}={}^{p}R_{ij}=0.
\end{equation}
Next in contrast to defining a new traceless stress tensor as in
\cite{Lysov:2011xx}, we insist to take the Brown-York stress tensor
as the fundamental variables and consider its fluctuations over the
background (To demonstrate that our scheme derives the same results
at the leading order of perturbations as in \cite{Lysov:2011xx}, we
present two examples in the appendix where the $p+1$-dimensional
hypersurface has a Minkowski limit). We expand the components of
Brown-York tensor in powers of $\lambda$
\begin{eqnarray}
    {t^\tau}_i&     = &0 +\lambda{ {t^\tau}_i}^{(1)} + \ldots\nonumber\\
    {t^\tau}_\tau&  = &0 + \lambda{{t^\tau}_\tau}^{(1)}+ \ldots\nonumber\\
    {t^i}_j&        = &{1\over{2\sqrt{r_c}}}{\delta^i}_j + \lambda{{t^i}_j}^{(1)} + \ldots\nonumber\\
    t\,\,&          = &{p\over{2\sqrt{r_c}}} + \lambda t^{(1)} +\ldots .
\end{eqnarray}
By definition in our formalism the relation $t=pK$ holds for
arbitrary order of the expansion and
$t^{(n)}={{t^{\tau}}_{\tau}}^{(n)}+{{t^i}_i}^{(n)}$. Substituting it
into the ``Hamiltonian constraint'' in Eq.(\ref{HC}) we find
\begin{equation}
    ({t^\tau}_\tau)^2 - {2\over{\lambda^2}}({t^\tau}_i)^2 +
    {t^i}_j{t^j}_i - {t^2\over{p}} =0.
\end{equation}
Note that all the indices here are lowered or raised with
$\gamma_{ab}$ or $\gamma^{ab}$. The leading order of the constraint
is $\lambda^{-2}$, which is automatically satisfied by the
background, while the sub-leading order of the expansion gives rise
to
\begin{equation}
    {{t^\tau}_\tau}^{(1)} =
    -2({{t^\tau}_i}^{(1)})({{t^\tau}_j}^{(1)})\gamma^{ij}.
\end{equation}
Next we turn to the Petrov type I condition. Firstly we choose the
vector fields as
\begin{eqnarray}
    \sqrt{2}\ell = \partial_0 - n,&
    \sqrt{2}k = -\partial_0 - n.\label{vf}
\end{eqnarray}
Then the condition becomes
\begin{equation}
    2C = C_{0i0j} +  C_{0ij(n)} +
    C_{0ji(n)} + C_{i(n)j(n)} = 0.
\end{equation}
With the use of Eq.(\ref{Weyl}), the Petrov type I condition can be
rewritten in terms of the Brown-York tensors as follows
\begin{eqnarray}
    {t^\tau}_\tau {t^i}_j +
    {2\over{\lambda^2}}\gamma^{ik}{t^\tau}_k{t^\tau}_j -
    2\lambda{t^i}_{j,\tau}  - {t^i}_k{t^k}_j -
    {2\over{\lambda}}\gamma^{ik}{t^\tau}_{(k,j)} +\nonumber\\
    {\delta^i}_j[{t\over{p}}({t\over{p}}-{t^\tau}_\tau)+2\lambda\partial_\tau{t\over{p}}]
    +{2\over{\lambda}}\gamma^{ik}{\Gamma^m}_{kj}{{t^\tau}_m} = 0.
\end{eqnarray}
First of all, after expanding in powers of $\lambda$, we find the
background satisfies this condition automatically at the order of
${1\over{\lambda^2}}:$
\begin{equation}
    -{1\over{4\lambda^2}}{\delta^i}_j +
    {1\over{4\lambda^2}}{\delta^i}_j = 0.
\end{equation}
The next non-vanishing order is $\lambda^0$, which gives rise to
the following equation
\begin{equation}
    {{t^i}_j}^{(1)} = 2\gamma^{ik}{{t^\tau}_k}^{(1)}{{t^\tau}_j}^{(1)} -
    2\gamma^{ik}{t^\tau}_{(k,j)}^{(1)} +
    {\delta^i}_j{t^{(1)}\over{p}} +
    2\gamma^{ik}{\Gamma^m}_{kj}{{t^\tau}_m}^{(1)}.\label{sp}
\end{equation}
Finally we come to the momentum constraints which is
\begin{equation}
    \nabla_a{t^a}_b = 0.
\end{equation}
The time component gives at leading order
\begin{equation}
    D_it^{\tau i(1)}=\partial_it^{\tau i(1)} + {\Gamma^i}_{ik}t^{\tau k(1)} =
    0,
\end{equation}
where $D_i\gamma_{jk}=0$. The space components at leading order
can be written as
\begin{equation}
    \partial_\tau {{t^\tau}_i}^{(1)} + D_k {{t^k}_i}^{(1)} = 0.
\end{equation}
Then plugging the solution to the Petrov type I condition in
equation (\ref{sp}) and identifying
\begin{eqnarray}
    {{t^\tau}_i}^{(1)} = {\upsilon_i\over{2}},&\,\,\,\,\,t^{(1)} =
    {p\over{2}} P,
\end{eqnarray}
we finally have the incompressible condition and the Navier-Stokes
equation in spatially curved spacetime as
\begin{equation}
    D_k\upsilon^k = 0,
\end{equation}
\begin{equation}
    \partial_\tau \upsilon_i + \upsilon^k D_k\upsilon_i + D_iP - (D^2\upsilon_i +
    {R^k}_i\upsilon_k) = 0.
\end{equation}
Since in this simple case ${}^pR_{ij}=0$, the last term in above
equation vanishes.

\section{Navier-Stokes equations in curved spacetime with non-vanishing $^pR_{ij}$}
In this section we explicitly construct a model with a non-vanishing
Ricci tensor $^pR_{ij}$. We assume that the metric of $p+2$
dimensional spacetime has the following form
\begin{equation}
ds^2_{p+2} = - f(r)dt^2 + 2dtdr +
e^{\rho(r,x^i)}\delta_{ij}dx^idx^j.\label{metric2}
\end{equation}
Now the spatial components of the metric is conformally flat, but
both $f$ and $\rho$ are functions of radial coordinate $r$. In
particular, we specify the function $f(r)$ has the following form
\begin{equation}
    f(r) = r(1 + a_1r + a_2 r^2 + \ldots)\label{fr}
\end{equation}
such that the Rindler horizon is fixed at $r=0$.

The hypersurface is located at $r=r_c$, then the induced metric
$\gamma_{ab}$ is
\begin{eqnarray}
    &ds^2&=-f(r_c)dt^2+e^{\rho(r_c,x^i)} \delta_{ij}dx^idx^j\nonumber\\
    &&=-{dx^0}^2+e^\rho \delta_{ij}dx^idx^j\equiv {-{1\over{\lambda^2}}d\tau^2+e^\rho \delta_{ij}dx^idx^j}.
\end{eqnarray}
Now it is straightforward to compute the components of the extrinsic
curvature, which are
\begin{eqnarray}
    K \,\,\,\,       &=&    {1\over{2\sqrt{f}}}\partial_r f + {1\over{2}}p\sqrt{f} \partial_r \rho \nonumber\\
    {K^\tau}_\tau    &=&    {1\over{2\sqrt{f}}}\partial_r f \nonumber\\
    {K^\tau}_i       &=&    0 \nonumber\\
    {K^i}_j          &=&    {1\over{2}}\sqrt{f}\partial_r
                            \rho{\delta^i}_j.
\end{eqnarray}
On the other hand, the intrinsic quantities of the hypersurface can
be obtained with the following components of connection:
\begin{eqnarray}
    {\Gamma^\tau}_{\tau \tau}  &=&  {\Gamma^\tau}_{\tau i} = {\Gamma^\tau}_{i j}={\Gamma^i}_{\tau \tau}={\Gamma^i}_{\tau j} = 0 \nonumber\\
    {\Gamma^i}_{jk}            &=&   {1\over{2}}
                                    ({\delta^i}_k\partial_{j} \rho + {\delta^i}_j {\partial_k} \rho
                                    - \delta^{im} \delta_{kj} \partial_m
                                    \rho).\label{cn}
\end{eqnarray}
Specifically, the components of Ricci tensor and the Ricci scalar
have the following form
\begin{eqnarray}
    ^{p+1}R_{\tau \tau}  &=&   ^{p+1}R_{\tau i}=0 \nonumber\\
    ^{p+1}R_{ij}\,       &=&   {2-p\over{2}}\partial_i\partial_j \rho -{1\over{2}}\delta_{ij}\delta^{km}\partial_k \partial_m\rho
                                + {p-2\over{4}}(\partial_i \rho)(\partial_j\rho) -
                               {p-2\over{4}}\delta_{ij}\delta^{km}(\partial_k\rho)(\partial_m\rho) \nonumber\\
     ^{p+1}R  \,\,\,     &=&   {(1-p)}\gamma^{ij}\partial_i\partial_j\rho
                               +{(1-p)(p-2)\over{4}}\gamma^{ij}(\partial_i\rho)(\partial_j\rho).\label{rrr}
\end{eqnarray}
Now we require that the metric in equation (\ref{metric2}) be a
solution to the Einstein vacuum equations in $p+2$ dimensions. After
a direct calculation, we find these equations to be
\begin{eqnarray}
    &\,&\partial_i \partial_r \rho                               =  0 \nonumber \\
    &\,&{\partial_r}^2\rho + {1\over{2}}{(\partial_r \rho)}^2    =  0  \nonumber \\
    &\,&{\partial_r}^2f + {1\over{2}}p(\partial_r f)(\partial_r \rho)=
    0\label{eve}
\end{eqnarray}
and
\begin{equation}
 ^{p+1}R_{ij} = \gamma_{ij}[{1\over{2}}(\partial_r f)(\partial_r \rho)
    + {pf\over{4}}{(\partial_r \rho)}^2 + {1\over{2}}f {\partial_r}^2
    \rho].\label{pR}
\end{equation}
 Observing the first equation in (\ref{eve}), we notice that its solution is the
 linear combination of arbitrary functions $F(x^i)$ and $G(r)$. This
 immediately leads to the fact that $^{p+1}R_{ij}$ is r-independent
 which can be seen from its definition in Eq.(\ref{rrr})\footnote{However $^{p+1}R$ is r-dependent, but only through the
 induced metric $\gamma^{ij}$.}. As a
 matter of fact, employing the equations in (\ref{eve}) we can show that the right hand
 side of Equation (\ref{pR}) is also r-independent by taking the partial derivative with respect to $r$.
Besides the trivial solution corresponding to the flat spacetime, we
can find the general solutions for $f(r)$ and $\rho(r,x^i)$ to be
\begin{eqnarray}
    f(r)\,\,\,\,    &=&      {{(r+c_1)}^{1-p}\over{1-p}}c_2 + c_3 \nonumber \\
    \rho(r,x^i)   &=&      F(x^i) + 2\ln{(r+c_1)},
\end{eqnarray}
where $c_1,c_2$ and $c_3$ are arbitrary constants, and the function
$F(x^i)$ is subject to the equation
\begin{equation}
     ^{p+1}R_{ij} =  c_3(p-1)e^{F(x^i)}\delta_{ij}.
\end{equation}
Obviously, non-trivial solutions to this equation exist, but we need
not to know their specific forms in our paper. Now requiring the
function $f(r)$ has the form as in (\ref{fr}), we can arrive at the
following solution after an appropriate choice of the constants
(here we set $c_1=c_2=1$ and $c_3={1\over{p-1}}$).
\begin{eqnarray}
    f(r)\,\,\,\,    &=&      {{{(r+1)}^{1-p}-1}\over{1-p}}  =   r - {p\over{2}}r^2 + {1\over{6}}(p+p^2)r^3 +
                         \ldots \nonumber \\
    \rho(r,x^i)   &=&     F(x^i) + 2\ln{(r+1)} = F(x^i) + 2r - r^2 +
                        {2\over{3}}r^3 + \ldots
\end{eqnarray}
Since we are mainly concerned with the behavior of the fluctuations
in
 the near horizon limit, we have expanded both functions $f(r)$ and $\rho(r,x^i)$
 in powers of $r$. As a consequence, the induced metric
 $\gamma_{ij}$ can also be expanded as
\begin{equation}
     \gamma_{ij} = e^{F(x^i)}\delta_{ij}(1+r)^2\equiv \gamma^{(0)}_{ij}+r \gamma^{(1)}_{ij}+r^2
     \gamma^{(2)}_{ij},
\end{equation}
where $\gamma^{(0)}_{ij}\equiv e^{F(x^i)}\delta_{ij}=
{}^{p+1}R_{ij}$. Moreover, we point out that the ``spatially
covariant derivative'' $D_i$ compatible with $\gamma_{ij}$ is also
compatible with $\gamma^{(n)}_{ij}$ since the connection is
r-independent, which can be seen by substituting the general form of
$\rho=F(x^i)+G(r)$ into Eq.(\ref{cn}).

Next we consider the effects of fluctuations of the extrinsic
curvature in a similar manner. The components of the Brown-York
stress tensor on $\Sigma_c$ are expanded as
\begin{eqnarray}
    {t^\tau}_i       &=&    0 + \lambda{{t^\tau}_i}^{(1)} + \ldots  \nonumber\\
    {t^\tau}_\tau    &=&    {1\over{2}}p\sqrt{f}\partial_r\rho+ \lambda{{t^\tau}_\tau}^{(1)} + \ldots  \nonumber\\
    {t^i}_j          &=&    ({1\over{2\sqrt{f}}}\partial_r f + {p-1\over{2}}\sqrt{f}\partial_r{\rho}){\delta^i}_j
                            +  \lambda {{t^i}_j}^{(1)} \ldots  \nonumber\\
    t\,\,            &=&    ({p\over{2\sqrt{f}}}\partial_r f + {1\over{2}}p^2\sqrt{f}\partial_r \rho)
                            + \lambda t^{(1)}+ \ldots.
\end{eqnarray}
When substituting these quantities into the Hamiltonian constraint
as well as the Petrov type I condition and expanding these equations
in powers of $\lambda$, we point out that the following quantities
should also be expanded since the location of the hypersurface $r_c$
is identified with $\lambda^2:$
\begin{eqnarray}
    \partial_r\rho |_{r_c} &=&      2 - 2r_c + 2{r_c}^2 + \ldots \nonumber \\
    \partial_r  f |_{r_c}  &=&      1 - pr_c + {1\over{2}}(p+p^2){r_c}^2 +
    \ldots \nonumber \\
    f^{1/2} |_{r_c}        &=&      {r_c}^{1/2} - {1\over{4}}p{r_c}^{3/2}
                               +{1\over{96}}(8p+5p^2){r_c}^{5/2} +
                               \ldots.
\end{eqnarray}
The ``Hamiltonian'' constraint is
\begin{equation}
    {({t^\tau}_\tau)}^2 - {2\over{\lambda^2}}({t^\tau}_i)^2 +
    {t^i}_j{t^j}_i - {t^2\over{p}} +\ ^{p+1}R = 0.
\end{equation}
The leading order of the expansion automatically vanishes with
\begin{equation}
    {p\over{4}} - {p\over{4}} = 0,
\end{equation}
while the non-trivial sub-leading order gives rise to
\begin{equation}
    {{t^\tau}_\tau}^{(1)} = -2\gamma^{ij(0)}{{t^\tau}_i}^{(1)}{{t^\tau}_j}^{(1)}.
\end{equation}
Next we turn to Petrov type I condition,
\begin{eqnarray}
    {t^\tau}_\tau {t^i}_j +
    {2\over{\lambda^2}}\gamma^{ik}{t^\tau}_k{t^\tau}_j -
    2\lambda{t^i}_{j,\tau}  - {t^i}_k{t^k}_j -
    {2\over{\lambda}}\gamma^{ik}{t^\tau}_{(k,j)} +\nonumber\\
    {\delta^i}_j[{t\over{p}}({t\over{p}}-{t^\tau}_\tau)+2\lambda\partial_\tau{t\over{p}}]
    +{2\over{\lambda}}\gamma^{ik}{\Gamma^m}_{kj}{{t^\tau}_m} - \gamma^{ik}R_{kj}= 0.
\end{eqnarray}
Similarly, taking the expansion we find the leading order is
automatically satisfied by the background quantities with
\begin{equation}
    -{1\over{4}}{\delta^i}_j + {1\over{4}}{\delta^i}_j = 0,
\end{equation}
and the non-trivial sub-leading order is $\lambda^0:$
\begin{equation}
    {{t^i}_j}^{(1)} = 2
    \gamma^{ik(0)}{{t^\tau}_k}^{(1)}{{t^\tau}_j}^{(1)}
    -2\gamma^{ik(0)}{t^\tau}_{(k,j)}^{(1)} +
    2\gamma^{ik(0)}{\Gamma^m}_{kj}{{t^\tau}_m}^{(1)}
    + {t^{(1)}\over{p}}{\delta^i}_j .
\end{equation}

Using the momentum constraint on the hypersurface and identifying
\begin{eqnarray}
    {{t^\tau}_i}^{(1)} = {1\over{2}}\upsilon_i,&\,\,\,\,P =
    {2\over{p}}t^{(1)},
\end{eqnarray}
we finally obtain the incompressible condition and the Navier-Stokes
equation in spatially curved background as
\begin{equation}
    {D_i}\upsilon^i = \partial_i \upsilon^i +
    {\Gamma^i}_{ik}\upsilon^k = 0
\end{equation}
\begin{equation}
   \partial_\tau \upsilon_i + D_i P + \upsilon^k {D_k}\upsilon_i -
   (D^{k}{D_k}\upsilon_i + {{R^m}_i}\upsilon_m) = 0,
\end{equation}
where we have used the fact that $D_i{{R^i}_j}=0$ since
$R_{ij}\propto \gamma_{ij}$.

\section{Summary and Discussions}
As a summary, we have generalized the framework in
\cite{Lysov:2011xx} by considering an embedding which may be
intrinsically curved. Directly employing the Brown-York stress
tensor as the fundamental variables of fluctuations, we explicitly
construct models with spatially curved embedding and demonstrate
that the incompressible Navier-Stokes equations can be derived for a
fluid moving on $\Sigma_c$ provided that the fluctuations are
subject to the Petrov type I condition as well as the ``Hamiltonian
constraint''. The fact that the Petrov type I condition can be
applied to a class of more general spacetime strongly implies that
this boundary condition would play a more important (and perhaps
fundamental) role in linking the Einstein equation and Navier-Stokes
equation, and this importance might be further disclosed by
manifestly proving the equivalence of imposing Petrov type I
condition with imposing the horizon regularity at least in the near
horizon limit.

Through the paper we need assume that the background is fixed
without time dependence such that the dynamics can be described by a
Navier-Stokes equation in a sense of a non-relativistic limit. In
the appendix the discussed models may have a dynamics for the
background, but constrained by the condition that it has a Minkowski
limit. It is an open question whether this framework could be
generalized to a general background which might be dynamical and
intrinsically curved.

In the end of this paper we propose that our current framework is
applicable to the Schwarzschild black holes\cite{Bredberg:2011xw}
and the spacetime in the presence of matter fields with a
cosmological constant. The investigation is under progress and will
be presented elsewhere\cite{HL}. We also expect that the higher
order expansions of the momentum constraint can be investigated in
the future.

\section*{Appendix: Two examples with a Minkowski limit}
\subsection{${ds^2}_{p+2}=-rdt^2+2dtdr+e^{\rho}\delta_{ij}dx^idx^j$}

Now we require that the hypersurface goes back to the flat
spacetime as $\lambda \rightarrow 0$, then the function $\rho$ can
be expanded as
\begin{equation}
\rho = 0 + \rho^{(1)}\lambda + \rho^{(2)}\lambda^2 +
\ldots\label{fb}.
\end{equation}
We stress that in this case the background need not to be fixed and
$\rho$ can be a general function of $(\tau,r,x^i)$. The hypersurface
is located at $r=r_c$ and the components of the connection
corresponding to the induced metric in $(\tau, x^i)$ coordinate
system read as
\begin{eqnarray}
    {\Gamma^\tau}_{\tau\tau}   &=&   {\Gamma^\tau}_{\tau i}={\Gamma^i}_{\tau\tau}=0\\
    {\Gamma^\tau}_{ij}         &=&   {1\over{2}} \lambda^2 e^\rho (\partial_{\tau}\rho) \delta_{ij} \\
    {\Gamma^i}_{\tau j}        &=&   {1\over{2}} {\delta^i}_j
    \partial_{\tau} \rho\\
    {\Gamma^i}_{jk}            &=&   {1\over{2}}
    ({\delta^i}_k\partial_{j} \rho + {\delta^i}_j {\partial_k} \rho
     - \delta^{im} \delta_{kj} \partial_m \rho).
\end{eqnarray}
Now it is straightforward to write down the ``Hamiltonian
constraint'' in terms of the Brown-York tress tensor and the
intrinsic curvature as
\begin{eqnarray}
    &&({t^\tau}_\tau)^2 -
    {2\over{\lambda^2}}({t^\tau}_i)^2 + {t^i}_j{t^j}_i - {t^2\over{p}}+
    p\lambda^2\partial_\tau^2\rho+{p(p+1)\over{4}}\lambda^2(\partial_\tau
    \rho)^2\nonumber\\
    &&+{(1-p)(p-2)\over{4}}e^{-\rho}\delta^{ij}(\partial_i
    \rho)(\partial_j\rho) +
    (1-p)e^{-\rho}\delta^{ij}\partial_i\partial_j \rho = 0.
\end{eqnarray}
Now consider the fluctuation effects of both the extrinsic curvature
and the background, we expand the variables in powers of $\lambda$
\begin{eqnarray}
    {t^\tau}_i&     = &0 + \lambda{ {t^\tau}_i}^{(1)} + \ldots\nonumber\\
    {t^\tau}_\tau&  = &{\lambda\over{2}}p\partial_\tau\rho + {p\over{2}}\sqrt{r_c}\partial_r\rho + \lambda{{t^\tau}_\tau}^{(1)} + \ldots\nonumber\\
    {t^i}_j&        = &({1\over{2\sqrt{r_c}}} + {p-1\over{2}}\lambda\partial_\tau\rho + {p-1\over{2}}\sqrt{r_c}\partial_r\rho){\delta^i}_j
                      + \lambda{{t^i}_j}^{(1)} + \ldots\nonumber\\
    t\,\,&          = &{p\over{2\sqrt{r_c}}} + {\lambda\over{2}}p^2\partial_\tau\rho + {p^2\over{2}}\sqrt{r_c}\partial_r\rho + \lambda t^{(1)} + \ldots\nonumber\\
    R&              = &0 + \lambda R^{(1)} + \ldots,
\end{eqnarray}
where $R^{(1)}$ can be found using Eq.(\ref{fb}). At the sub-leading
order of the hamiltonian constraint we have
\begin{equation}
    {{t^\tau}_\tau}^{(1)} =
    -2{{t^\tau}_i}^{(1)}{{t^\tau}_j}^{(1)}\delta^{ij}.
\end{equation}
In a parallel way the Petrov type I condition leads to the following
form
\begin{eqnarray}
    {t^\tau}_\tau {t^i}_j +
        {2\over{\lambda^2}}\gamma^{ik}{t^\tau}_k{t^\tau}_j -
        2\lambda{t^i}_{j,\tau}  - {t^i}_k{t^k}_j -
        {2\over{\lambda}}\gamma^{ik}{t^\tau}_{(k,j)} +
        {\delta^i}_j[{t\over{p}}({t\over{p}}-{t^\tau}_\tau)+2\lambda\partial_\tau{t\over{p}}] + \lambda [ -{t^i}_j\partial_\tau\rho \nonumber\\
    + {t^\tau}_\tau{\delta^i}_j\partial_\tau\rho + {1\over{\lambda^2}} ( {t^\tau}_j\gamma^{ik}\partial_k\rho
        + t^{\tau i}\partial_j\rho - {\delta^i}_j\gamma^{km}{t^\tau}_k\partial_m\rho ) ]
       -[{1+p\over{4}}\lambda^2{\delta^i}_j{(\partial_\tau \rho)}^2 +  \lambda^2{\delta^i}_j{\partial_\tau}^2\rho\,\,\,\, \nonumber\\
    + {p-2\over{4}}\gamma^{ik}(\partial_k \rho)(\partial_j
        \rho) + {2-p\over{2}}\gamma^{ik}\partial_k \partial_j\rho +
        {2-p\over{4}}{\delta^i}_j\gamma^{km}(\partial_k\rho)(\partial_m\rho)
        -{1\over{2}}{\delta^i}_j\gamma^{km}\partial_k\partial_m\rho ]  =
        0.\,\,\,\,\,
\end{eqnarray}
The sub-leading term of ${t^i}_j$ is
\begin{equation}
    {{t^i}_j}^{(1)} = 2\delta^{ik}{{t^\tau}_k}^{(1)}{{t^\tau}_j}^{(1)} -
    2\delta^{ik}{t^\tau}_{(k,j)}^{(1)} +
    {\delta^i}_j{t^{(1)}\over{p}}.
\end{equation}
Finally, from the momentum constraint
\begin{equation}
    \nabla_a{t^a}_b = 0,
\end{equation}
we have the time component as
\begin{equation}
    \partial_\tau{t^\tau}_\tau -
    {1\over{\lambda^2}}\partial_it^{\tau i} +
    {p+1\over{2}}{t^\tau}_\tau\partial_\tau \rho  -
    {p\over{2\lambda^2}}t^{\tau k}\partial_k\rho  -
    {t\over{2}}\partial_\tau \rho = 0.
\end{equation}
The leading order of the expansion gives
\begin{equation}
    \partial_i t^{\tau i(1)} = 0.
\end{equation}
The space components of the constraint is
\begin{equation}
    (\partial_\tau{t^\tau}_i - {1\over{2}}{t^\tau}_i\partial_\tau \rho) + \partial_k{t^k}_i
    + {1+p\over{2}}{t^\tau}_i\partial_\tau\rho +
    {p\over{2}}{t^k}_i\partial_k\rho -
    {1\over{2}}(t-{t^\tau}_\tau)\partial_i\rho = 0.
\end{equation}
We find the leading order of the expansion is automatically
satisfied with a form as
\begin{equation}
    {p\over{4}}\partial_i{\rho}^{(1)} - {p\over{4}}\partial_i{\rho}^{(1)} =
    0,
\end{equation}
while the sub-leading order leads to
\begin{equation}
    \partial_\tau{{t^\tau}_i}^{(1)} + 2t^{\tau k(1)}\partial_k{{t^\tau}_i}^{(1)} - \partial^2{{t^\tau}_i}^{(1)} + \partial_i{t^{(1)}\over{p}} =
    0.
\end{equation}
Identifying
\begin{eqnarray}
    {{t^\tau}_i}^{(1)} = {\upsilon_i\over{2}}, &\,\,\,\,\,t^{(1)} =
    {p\over{2}} P,
\end{eqnarray}
we obtain a Navier-Stokes equation with incompressible condition for
a fluid in p-dimensional flat space.
\begin{equation}
    \partial_i\upsilon^i = 0,
\end{equation}
\begin{equation}
    \partial_\tau\upsilon_i + \upsilon^k\partial_k\upsilon_i - \partial^2\upsilon_i + \partial_iP =
    0.
\end{equation}
From above expansion, we notice that the contribution from the
fluctuations of the background are higher order such that the
leading order of the solutions to the Hamiltonian constraint and the
Petrov I condition are the same as those in flat embedding with a
fixed background, which of course is not surprising since the
spacetime is subject to the condition of having a Minkowski limit.
However, we would like to point out that the fluctuations of the
background certainly will provide corrections to the higher order
expansions, which should be different from the results with a fixed
background£¬ and we leave this issue for study in future.

\subsection{$ {ds^2}_{p+2} = -re^\rho dt^2 + 2drdt + e^\rho\delta_{ij}dx^idx^j $}
The induced metric on the hypersurface $r=r_c$ is conformally flat,
even the effects of fluctuations are taken into account. Therefore,
we may set the trace of the Brown-York tensor to be a constant
without fluctuations, as suggested in \cite{Lysov:2011xx}. The
components of the connection in $p+1$ dimensional spacetime are
\begin{eqnarray}
    {\Gamma^\tau}_{\tau\tau} = {1\over{2}}\partial_\tau\rho ,\,\,\,\,\,
        {\Gamma^\tau}_{\tau i} = {1\over{2}}\partial_i\rho , \,\,\,\,\,{\Gamma^\tau}_{ij} = {1\over{2}} \lambda^2\delta_{ij}\partial_{\tau}\rho ,\ \ \  {\Gamma^i}_{\tau\tau} =
        {1\over{2\lambda^2}}\delta^{ij}\partial_j \rho
        , \nonumber \\
    {\Gamma^i}_{\tau j}={1\over{2}} {\delta^i}_j \partial_{\tau} \rho ,\
        \ \ \ \ {\Gamma^i}_{jk}={1\over{2}} ({\delta^i}_k
        \partial_{j} \rho +
        {\delta^i}_j {\partial_k} \rho
        - \delta^{im} \delta_{kj} \partial_m
        \rho).\,\,\,\,\,\,\,\,\,\,\,\,\,\,\,
\end{eqnarray}
The ``Hamiltonian constraint'' has the form
\begin{eqnarray}
    & & ({t^\tau}_\tau)^2 -
    {2\over{\lambda^2}}{t^\tau}_i{t^\tau}_j\delta^{ij} + {t^i}_j{t^j}_i - {{t}^2\over{p}}
    +
    e^{-\rho}(p{\lambda^2}{\partial_\tau}^2\rho-p\delta^{ij}\partial_i\partial_j\rho)
      \nonumber\\
   & &+{{p-p^2}\over{4}}e^{-\rho}[-{\lambda^2}(\partial_\tau\rho)^2+\delta^{ij}(\partial_i\rho)(\partial_j\rho)]
    = 0.
\end{eqnarray}
It is easy to check that the leading order are satisfied by the
quantities of the background, and the sub-leading order vanishes
with
\begin{equation}
    {\rho^{(1)}\over{4}}p - {\rho^{(1)}\over{4}}p = 0.
\end{equation}
The subsequent order is $\lambda^{0}$ which gives
\begin{equation}
    {{t^\tau}_\tau}^{(1)} = -2{{t^\tau}_i}^{(1)}{{t^\tau}_j}^{(1)}\delta^{ij}.
\end{equation}
While for Petrov type I condition we introduce the following vector
fields
\begin{eqnarray}
    m_i = e^{-{\rho\over{2}}}\partial_i,&\sqrt{2}\ell = e^{-{\rho\over{2}}}\partial_0 - n,&
    \sqrt{2}k = -e^{-{\rho\over{2}}}\partial_0 - n,
\end{eqnarray}
then the condition is written as
\begin{equation}
    2C = e^{-\rho}C_{0i0j} + e^{-{\rho\over{2}}} C_{0ij(n)} +
    e^{-{\rho\over{2}}}C_{0ji(n)} + C_{i(n)j(n)} = 0.
\end{equation}
In terms of the extrinsic curvature and the components of the
induced metric, it can be rewritten as
\begin{eqnarray}
    &&{t^\tau}_\tau {t^i}_j +
        {2\over{\lambda^2}}\delta^{ik}{t^\tau}_k{t^\tau}_j -
        2\lambda e^{-{\rho\over{2}}}{t^i}_{j,\tau} - {t^i}_k{t^k}_j  +
        {\delta^i}_j[{t\over{p}}({t\over{p}}-{t^\tau}_\tau)+2\lambda
        e^{-{\rho\over{2}}}\partial_\tau{t\over{p}} ]  \nonumber\\
    &&- {2\over{\lambda}}e^{-{\rho\over{2}}}\delta^{ki}{t^\tau}_{(k,j)}
        +[ -\lambda^2e^{-{\rho}}{\delta^i}_j{\partial_\tau}^2\rho +
        {p\over{2}}\gamma^{ik}\partial_k\partial_j\rho +
        {p\over{4}}{\delta^i}_j\gamma^{km}(\partial_k\rho)(\partial_m\rho)  \nonumber\\
    &&- {p\over{4}}\gamma^{ik}(\partial_k\rho)(\partial_j\rho)
        -{p-1\over{4}}\lambda^2 e^{-\rho}{\delta^i}_j(\partial_\tau\rho)^2 +
        {1\over{2}}{\delta^i}_j\gamma^{km}\partial_k\partial_m\rho ]  \nonumber\\
    &&+ e^{-{\rho\over{2}}}\lambda [( -{t^i}_j + {\delta^i}_j{t^\tau}_\tau ) \partial_\tau\rho +
        {1\over{\lambda^2}} ( {1\over{2}}{t^\tau}_j\delta^{ik}\partial_k\rho - e^{\rho}{\delta^i}_j
        t^{\tau k}\partial_k\rho + {1\over{2}}e^\rho t^{\tau i}\partial_j\rho )] =0.
\end{eqnarray}
Although in general this equation is rather complicated, its leading
expansions in power of $\lambda$ become very simple when the metric
of the background has a Minkowski limit as described in (\ref{fb}).
It turns out that the leading order of ${t^i}_j$ still has the form
\begin{equation}
    {{t^i}_j}^{(1)} = {t^{(1)}\over{p}}{\delta^i}_j -
    2\delta^{ik}{t^\tau}_{(k,j)}^{(1)} + 2t^{\tau i
    (1)}{{t^\tau}_j}^{(1)}.
\end{equation}
The time component of the momentum constraint is
\begin{equation}
    \partial_\tau {t^\tau}_\tau - {1\over{\lambda^2}}\partial_i(t^{\tau
    i}e^{\rho}) -
    {1\over{2\lambda^2}}\delta^{ik}{t^\tau}_k\partial_i\rho +
    {1+p\over{2}}{t^\tau}_\tau\partial_\tau\rho -
    {1\over{2}}t\partial_\tau\rho - {p\over{2\lambda^2}}e^{\rho} t^{\tau
    k}\partial_k\rho = 0.
\end{equation}
Its leading order gives rise to
\begin{equation}
    \partial_it^{\tau i(1)} = 0.
\end{equation}
While the space components of the constraint is
\begin{equation}
    \partial_\tau {t^\tau}_i + \partial_k {t^k}_i +
    {p+1\over{2}}{t^k}_i\partial_k\rho +
    {p+1\over{2}}{t^\tau}_i\partial_\tau\rho - {t\over{2}}\partial_i\rho
    = 0.
\end{equation}
Its leading and sub-leading orders of the expansion respectively
lead to
\begin{equation}
    \partial_i\rho^{(1)} = 0,
\end{equation}
and
\begin{equation}
    \partial_\tau{{t^\tau}_i}^{(1)} + {1\over{p}}\partial_it^{(1)} +
    2t^{\tau k{(1)}}\partial_k{{t^\tau}_i}^{(1)} +
    {1\over{2}}\partial_i\rho^{(2)} - \partial^2{{t^\tau}_i}^{(1)} =
    0.
\end{equation}
Identifying
\begin{eqnarray}
    {{t^\tau}_i}^{(1)} = {\upsilon_i\over{2}},&\,\,\,\,\,& \rho^{(2)} = P
\end{eqnarray}
and taking $t^{(1)}=0$, we arrive at the Navier-Stokes equation with
incompressible condition.
\begin{equation}
    \partial_i\upsilon^i = 0,
\end{equation}
\begin{equation}
    \partial_\tau\upsilon_i + \upsilon^k\partial_k\upsilon_i -
    \partial^2\upsilon_i + \partial_iP = 0.
\end{equation}

\begin{acknowledgments}
We are grateful to Chao-Guang Huang and Hongbao Zhang for
discussion and correspondence. Y.Ling would like to thank Prof.
Xiaoning Wu for invitation and the AMSS, CAS for hospitality
during his visit, where this work was initiated. T. Huang, Y. Ling
and W.Pan are partly supported by NSFC(10875057), Fok Ying Tung
Education Foundation(No. 111008), the key project of Chinese
Ministry of Education(No.208072), Jiangxi young
scientists(JingGang Star) program and 555 talent project of
Jiangxi Province. Y. Tian and X. Wu are partly supported by NSFC
(Nos. 10705048, 10731080 and 11075206) and the President Fund of
GUCAS. We also acknowledge the support by the Program for
Innovative Research Team of Nanchang University.
\end{acknowledgments}

\end{document}